\documentclass[%
 prx,
 jmp,%
 amsmath,amssymb,
 floatfix,
reprint,%
 showpacs,%
]{revtex4-2}

\usepackage{comment}
\usepackage{amssymb}
\usepackage{amsmath}
\usepackage{mathtools}
\usepackage{relsize}
\usepackage{array}
\usepackage{bm}
\usepackage{graphicx}
\usepackage{bigints}
\usepackage{ifthen}
\usepackage{multirow}
\usepackage{threeparttable}
\usepackage{rotating}
\usepackage{hyperref}
\usepackage{xcolor}
\usepackage{empheq}
\usepackage[normalem]{ulem}
\usepackage{media9}

\usepackage[draft,footnote,marginclue,innerlayout=inline]{fixme}
\fxloadlayouts{marginnote}
\fxsetup{theme=color,mode=multiuser}
\FXRegisterAuthor{V}{anV}{{\color{red}{VG}}}

\DeclareGraphicsExtensions{.pdf,.png,.eps}
\graphicspath{{figs/}{./}}
\usepackage{cleveref}
\crefname{equation}{}{}
\Crefname{equation}{}{}
\crefname{figure}{Fig.~\!\!\!}{Figs.~\!\!\!}
\Crefname{figure}{Fig.~\!\!\!}{Figs.~\!\!\!}

\usepackage{tikz}
\usetikzlibrary{intersections,arrows,decorations,decorations.pathmorphing,calc}
\usepackage{pgfplots}
\pgfplotsset{compat=newest}
\pgfplotsset{plot coordinates/math parser=false}

\makeatletter
\let\jnl@style=\rmfamily
\def\ref@jnl#1{{\jnl@style#1}}
\@ifundefined{aap}{\newcommand\aap{\ref@jnl{A\&A}}}{}%
\@ifundefined{apj}{\newcommand\apj{\ref@jnl{ApJ}}}{}%
\@ifundefined{apjl}{\newcommand\apjl{\ref@jnl{Astrophys.~J.~Lett.}}}{}%
\@ifundefined{ssr}{\newcommand\ssr{\ref@jnl{Space~Sci.~Rev.}}}{}%
\@ifundefined{jgr}{\newcommand\jgr{\ref@jnl{J.~Geophys.~Res.}}}{}%
\@ifundefined{jatp}{\newcommand\jatp{\ref@jnl{J.~Atmos.~Terr.~Phys.}}}{}%
\@ifundefined{solphys}{\newcommand\solphys{\ref@jnl{Sol.~Phys.}}}{}%
\@ifundefined{pp}{\newcommand\pp{\ref@jnl{Phys.~Plasmas}}}{}%
\@ifundefined{mnras}{\newcommand\mnras{\ref@jnl{MNRAS}}}{}%
\@ifundefined{apjs}{\newcommand\apjs{\ref@jnl{ApJS}}}{}%
\@ifundefined{azh}{\newcommand\azh{\ref@jnl{AZh}}}{}%
\catcode`\@=11

\def\beq#1\eeq{\begin{equation}#1\end{equation}}
\def\bes#1\ees{\begin{subequations}#1\end{subequations}}
\def\bea#1\eea{\begin{align}#1\end{align}}
\def\n{\nonumber\\}

\newcommand{\D}[2]{\frac{\partial #1}{\partial #2}}

\newcommand{\mvec}[1]{\bm{#1}}

\usepackage{scalerel}

\newcommand{\der}[2]{\frac{d {#1}}{d {#2}}}

\begin{document}

\title{Universal mechanism of low--frequency brain rhythm formation
  through nonlinear coupling of high--frequency spiking--like
  activity}

\author{Vitaly L. Galinsky}
\email{vit@ucsd.edu}
\affiliation{Center for Scientific Computation in Imaging,
University of California at San Diego, La Jolla, CA 92037-0854, USA}
\affiliation{Department of ECE, University of California, San Diego,
  La Jolla, CA 92093-0407, USA}
\author{Lawrence R. Frank}
\email{lfrank@ucsd.edu}
\affiliation{Center for Scientific Computation in Imaging,
University of California at San Diego, La Jolla, CA 92037-0854, USA}
\affiliation{
Center for Functional MRI,
University of California at San Diego, La Jolla, CA 92037-0677, USA}

\date{\today}

\begin{abstract}
A universal mechanism of emergence of synchronized low frequency brain
wave field activity is presented as a result of nonlinear coupling
with flat frequency neuronal forcing.  The mechanism utilizes a unique
dispersion properties of weakly--evanescent wave--like brain surface
modes that are predicted to exist within a inhomogeneous and
anisotropic physical brain tissue model. These surface modes
are able to propagate in thin inhomogeneous layers with frequencies
that are inverse proportional to wave numbers. The resonant and
non--resonant terms of nonlinear coupling between multiple modes
produce both synchronous spiking--like high frequency wave activity as
well as low frequency wave rhythms. The relatively narrow localized
frequency response of the non--resonant coupling can be expressed by
terms similar to phase coupling in oscillatory systems.  Numerical
simulation of forced multiple mode dynamics shows as forcing increases
a transition from damped to oscillatory regime that is then silenced
off as over excitation is reached.  The resonant nonlinear coupling
results in emergence of low frequency rhythms with frequencies that
are several orders of magnitude below the linear frequencies of modes
taking part in the coupling.
\end{abstract}

\maketitle

The abundance of oscillatory patterns across a
wide range of spatial and temporal scales of brain electromagnetic
activity makes a question of their interaction an important
issue that has been widely discussed in the literature
\citep{buzsaki2006rhythms,*Gerstner:2014:NDS:2635959}. The
standard approach involves representing the
brain as a large network of coupled oscillators
\citep{2000PhyD..144...62F,*2002PhyD..163..191G} and using this as a
testbed for the study of network wave propagation,
mechanisms of synchrony, possibly deriving some mean field equations
and properties, etc.  However, such models are necessarily
descriptive and their relationship to actual physical properties of
either to actual brain tissue properties or the electromagnetic
waves they support is tenuous.

In this paper we employ a different approach that uses properties of
brain waves in realistic brain tissue types and
architectures derived in a general form from relatively basic
physical principles \citep{bwl}. We will demonstrate that
the peculiar inverse proportionality of the wave linear dispersion
found in \citep{bwl} combined with nonlinear resonant and
non--resonant coupling of multiple wave modes produces a
remarkably feature rich nonlinear system that is able to reproduce
many seemingly unrelated regimes that have been observed
experimentallly throughout a wide range of
scales of brain activity. The different regimes
include high frequency spiking--like activity occurring near
the critical point of the equation that integrates
multiple non--resonant wave modes and low frequency oscillations that
emerge when weak resonant coupling is present in the vicinity of the
critical point. The strongly nonlinear regime exists sufficiently
close to the critical point where the solution bifurcates from
oscillatory to non--oscillatory behavior. The weak resonant coupling
then demonstrates a mechanism that constantly moves the
system back and forth from subcritical to supercritial domains turning
the spiking on and off with low frequency quasiperiodicity.

In order to describe this complex behavior we show
for the first time that the inverse proportionality of frequency and
wavenumber in brain wave dispersion relation permits the
characterization of a limiting form for the signals in terms
of a large number of wavemodes as a
summation of non-resonant wave harmonics, thus allowing
a closed analytical form of nonlinear equation that
integrates and includes the collective non-resonant input from
multiple wave modes. Following the ideas of wave turbulence
\citep{book:971420,*book:787941} we also show that the resonant
coupling between those high frequency nonlinear wave modes can provide
an effective universal mechanism for the emergence of low frequency
wave rhythms.

Following \citep{bwl} we will use Maxwell equations in a medium for 
description of brain electromagnetic activity
\makeatletter%
\if@twocolumn%
  \begin{align}
  \nabla\cdot\mvec{D} &= \rho,\quad
  \nabla\times\mvec{H} = \mvec{J} + \D{\mvec{D}}{t}\quad \Rightarrow \quad
  \D{\rho}{t} &+ \mvec{\nabla}\cdot\mvec{J} = 0.\nonumber
  \end{align}
\else
  \begin{align}
  \nabla\cdot\mvec{D} &= \rho,\quad
  \nabla\times\mvec{H} = \mvec{J} + \D{\mvec{D}}{t}\quad \Rightarrow \quad
  \D{\rho}{t} + \mvec{\nabla}\cdot\mvec{J} = 0.\nonumber
  \end{align}
\fi
\makeatother

Using the electrostatic potential $\mvec{E}=-\nabla \phi$, Ohm's law
$\mvec{J}=\mvec{\sigma}\cdot\mvec{E}$ (where
$\mvec{\sigma}\equiv\{\sigma_{ij}\}$ is an anisotropic conductivity
tensor), a linear electrostatic property for brain tissue
$\mvec{D}=\varepsilon\mvec{E}$, assuming that the permittivity is a
``good'' function (i.e. it does not go to zero or infinity
anywhere) and taking the change of
variables $\partial x \to \varepsilon \partial x^\prime$, the charge
continuity equation for the spatial-temporal evolution of the
potential $\phi$ can be written in terms of a permittivity scaled
conductivity tensor $\mvec{\Sigma}=\{\sigma_{ij}/\varepsilon\}$ as

\begin{align}
\label{eq:phiSigma}
\D{}{t} \left(\nabla^2 \phi \right) &=
-\mvec{\nabla}\cdot\mvec{\Sigma}\cdot \nabla\phi + \mathcal{F}, 
\end{align}
where we have included a possible external source (or forcing) term
$\mathcal{F}$.  For brain fiber tissues the conductivity tensor
$\mvec{\Sigma}$ might have significantly larger values along the fiber
direction than across them. Taking into account an inhomogeneity of
the conductivity tensor $\mvec{\Sigma}$ this system shows existence of
weakly--evanescent wave modes \citep{bwl}.  We assume for simplicity
a two dimensional symmetric form of the conductivity tensor
with constant diagonal terms $\Sigma_{xx}$ and $\Sigma_{yy}$ (where
$\Sigma_{yy}$ is along the fibers conductivity, $\Sigma_{xx} <
\Sigma_{yy}$) and position dependent off--diagonal terms $\Sigma_{xy}$
that are changing linearly with $y$ through a relatively narrow layer
at the boundary so that the conductivity gradient exists only inside
this layer and is directed along the $y$ axis. We will only be
interested in a one dimensional solution for the potential $\phi(x)$
located in this thin layer of inhomogeneity that can be described by
the reduced equation
\begin{align}
\label{eq:phiTensor}
\partial_t \partial_x^2 \phi &+ \gamma_d\partial_x^2\phi +
\Omega\partial_x\phi=\mathcal{F},
\end{align}
where $\gamma_d = \Sigma_{xx}$ and $\Omega =\partial_y \Sigma_{xy}$.

The source term $\mathcal{F}$ can be assumed to have a frequency
independent forcing part with a linear growth rate
$\gamma_e$ representing some averaged input from random spiking
activity and an additional term that describes the nonlinear
amplitude/phase coupling of the firing rate to the wave field itself
\cite{pmid11832222,*pmid3720881,*pmid1521610,*pmid21414915},
\begin{align}
\label{eq:source}
\mathcal{F} = -\gamma_e\phi - \mathcal{N}(\phi).
\end{align}

The solution $\phi$ can be sought as a Fourier integral expansion 
\begin{align}
\label{eq:series}
\phi(x,t) = \int
a(k,t) e^{i\left(k x + \omega_{k}t\right)}dk + c.c.,
\end{align}
for wave modes with frequencies $\omega_k$ and wave numbers $k$ (where 
``c.c.'' denotes complex conjugate),
that results in a set of coupled equations for time dependent complex 
amplitudes $a_k(t)\equiv a(k,t)$
\begin{align}
\label{eq:ak}
\der{a_k}{t} &= \left(\frac{\gamma_e}{k^2} - \gamma_d\right) a_k +
\frac{1}{k^2}\mathcal{N}_k, 
\end{align}
where the wave mode frequencies are inversely proportional to the wave number
\begin{align}
\label{eq:disp}
\omega_k = \Omega/k,  \qquad |k| > k_{0} = 2 \pi/L,
\end{align}
and 
\begin{align}
\label{eq:NAi}
\mathcal{N}_k = \frac{1}{2\pi}\int \mathcal{N}(\phi) e^{-i\left(k x +
  \omega_{k}t\right)}dx. 
\end{align}

The nonlinear terms $\mathcal{N}_k$ will include a sum of inputs from
multiple waves, i.e., $k = \sum_{i}^{n} k_{n}$ where $n$ is
the order of the non-linearity. Those resonant conditions will give rise to
coupling terms that includes various combinations of
$\exp(i(\omega_k - \sum_{i}^{n}\pm\omega_{k_i})t)$.
Additional requirements for frequency resonances
($\omega_k= \sum_{i}^{n}\pm\omega_{k_i}$) produces wave
turbulence-like \citep{book:971420,*book:787941} selection rules for
the coupling terms that are
similar to phase coupling terms in a ring of connected oscillators
\citep{Kuramoto2002,*Kuramoto2003}.

For waves having typical dispersion properties, that is with the
frequencies directly proportional to the wave numbers ($\omega_k\sim
k^\alpha$ $\alpha>0$), the maximum oscillatory frequency is increasing
and going to infinity with increase of wave numbers. In this case the
nonlinear terms produce a direct cascade of wave energy
\citep{book:971420,*book:787941} constantly generating larger and
larger frequencies. For the inversely proportional wave dispersion,
like the waves considered in this paper, the wave energy will be
cascaded into smaller frequencies, thus
providing a natural mechanism for
synchronization of high frequency spiking input and emergence of low
frequency rhythms.

This model is also able to characterize another important
phenomenon whose existence is supported by an abundance of
experimental data  --
feedback between field potential and firing rate
\citep{pmid11832222,*pmid3720881,*pmid21414915,*pmid1521610}. The
feedback can be represented through nonlinear coupling.  This
will be demonstrated using the simplest
quadratic form $\mathcal{N}(\phi)=\phi(x,t)^2$ for the coupling
which can arise through many different processes.  More
complex feedback can be generated by higher order coupling terms of
course but that discussion is beyond the scope of this current
paper. The quadratic form of coupling results in
\begin{align}
\label{eq:Nk}
\mathcal{N}_k 
&=\!
\int\!
\left[
\delta(k\pm k' \pm k'')e^{-i(\omega_k-\omega_{k'}-\omega_{k''})t}a_{k'}a_{k''}
\right]
dk' dk''.
\end{align}
Using symmetry conditions $a_{-k} = a^*_k$ and $\omega_{-k} = -
\omega_k$ (a consequence of $\phi^*(x)=\phi(x)$) this can be rewritten
as
\begin{align}
\label{eq:Nkn}
\mathcal{N}_k 
=\!
\int\!
\left[\vphantom{\int}\right.&\left.
e^{-i(\omega_k-\omega_{k'}-\omega_{k-k'})t}a_{k'}a_{k-k'}
\right.\n+&\left.
e^{-i(\omega_k-\omega_{k'}+\omega_{k-k'})t}a_{k'}a^*_{k-k'}
\right.\\+&\left.
e^{-i(\omega_k-\omega_{k'}-\omega_{k+k'})t}a_{k'}a_{k+k'}
\right.\n+&\left.
e^{-i(\omega_k-\omega_{k'}+\omega_{k+k'})t}a_{k'}a^*_{k+k'}
\vphantom{\int}
\right]
dk' .\nonumber
\end{align}
The corresponding conditions for the frequency resonances $\omega_{k}
= \pm \omega_{k'} \pm \omega_{k''}$ allow the expression of
the nonlinear resonant coupling $\mathcal{N}^{R}_k$ by
extraction of only the relevant terms as
\begin{align}
\label{eq:Nko}
 \mathcal{N}^{R}_k 
&\sim  \left[
 a_{k_{-2}}a^*_{k_{-1}}
 +
 a_{k_{-1}}a_{k_{1}}
 +
 a^*_{k_{1}}a_{k_{2}} \right],
\end{align}
where only three out of four wave number resonances appear, as the
resonance $k + k' + k'' = 0$ is not possible
\citep{book:971420,*book:787941}, and $k_{-2} = k(3-\sqrt{5})/2$,
$k_{-1} = k(1-\sqrt{5})/2$, $k_1=k(-1-\sqrt{5})/2$, and
$k_2=k(3+\sqrt{5})/2$, are the real solutions of quadratic equations
$1/k \pm 1/k' \pm 1/|k-k'|=0$.

An important addition to these coupling terms arises from the inverse
proportionality of frequency and wave number in the dispersion
relation \cref{eq:disp}. The difference of frequencies of nonlinear
non--resonant harmonics is decreasing and going to zero with
increasing wave number, thus effectively
allowing a closed form expression for the limit of
$k\rightarrow\infty$, an effect that is absent for coupling of waves
with directly proportional dispersion. To illustrate this,
we will estimate the non--resonant nonlinear input
$\mathcal{N}^{nR}_{k_0}$ to the $k_0$ wave mode.
\begin{align}
\label{eq:Nknr}
\mathcal{N}^{nR}_{k_0}
=\!
\int\!
\left[\vphantom{\int}\right.&\left.
e^{-i\delta\omega_1(k)t}a_{k}a^*_{k-k_0}
\right.\n+&\left.
e^{-i\delta\omega_2(k)t}a_{k}a_{k-k_0}
\right.\\+&\left.
e^{-i\delta\omega_3(k)t}a_{k}a^*_{k+k_0}
\right.\n+&\left.
e^{-i\delta\omega_4(k)t}a_{k}a_{k+k_0}
\vphantom{\int}
\right]
dk,\nonumber
\end{align}
where 
\begin{align}
\delta\omega_1(k) &= \omega_{k_0}-\omega_{k}+\omega_{k-k_0},\n
\delta\omega_2(k) &= \omega_{k_0}-\omega_{k}-\omega_{k-k_0},\n
\delta\omega_3(k) &= \omega_{k_0}-\omega_{k}+\omega_{k+k_0},\n
\delta\omega_4(k) &= \omega_{k_0}-\omega_{k}-\omega_{k+k_0}.\nonumber
\end{align}

The approximate expression for the forced oscillations solution can be
obtained assuming that all the forcing input originates from the
scales of $k=k_0$ (the forcing term $\gamma_e/k^2$ in
\cref{eq:ak} is largest when $k=k_0$). Therefore we will derive the
non--resonant input term $\mathcal{N}^{nR}_{k_0}$ only for $k=k_0$,
thus neglecting nonlinear and damping terms for any $k>k_0$ (more
correctly, for any $k$ that are not in resonance with $k_0$ or for
$k>k_{2} = k_0(3+\sqrt{5}/2)\approx 2.618 k_0)$.  At the limit
$k\rightarrow\infty$ all frequency deltas $\delta\omega_{1-4}(k)
\rightarrow \omega_{k_0} \equiv \omega_0$ and $k-k_0\approx k$, hence
approximately we can estimate the non--resonant term
$\mathcal{N}^{nR}_{k_0}$ as
\begin{align}
\label{eq:nra}
\mathcal{N}^{nR}_{k_0} &\approx 
2 e^{-i\omega_{k_0}t}
\int\left[ a_k a^*_k + a_k^2\right] dk \n
&\approx
e^{-i\omega_{k_0}t}
\int\left[ a_k + a^*_k\right]^2 dk.
\end{align}

To estimate forced oscillations terms required in evaluation of the
integral \cref{eq:nra}, one can write from \cref{eq:ak}, \cref{eq:Nkn}
and \cref{eq:Nknr} that
\begin{align}
\label{eq:aknr}
\der{a_k}{t}
=\frac{1}{k^2}\left[
\vphantom{e^{i\delta\omega_1(k) t}a_{k_0}a_{k-k_0}}
\right.
&\left.
e^{i\delta\omega_1(k) t}a_{k_0}a_{k-k_0} 
\right.\n
+&\left.
e^{i\delta\omega_2(k) t}a_{k_0}a^*_{k-k_0}
\right.\n
+&\left.
e^{i\delta\omega_3(k) t}a_{k_0}a_{k+k_0}
\right.\n
+&\left.
e^{i\delta\omega_4(k) t}a_{k_0}a^*_{k+k_0}
\right].
\end{align}

Looking again for an approximate large $k$ solution ($a_{k-k_0}
\approx a_k$ for $k\gg k_0$) and keeping only terms that include
$a_{k_0}$ (assuming that the amplitude $a_{k_0}$ is small and can be
considered constant relative to any of the $\delta\omega(k)$ terms),
we can approximately write that
\begin{align}
\label{eq:aknrsol}
a_k
&
\approx \frac{-i a_{k_0}}{k^2}
\sum_{j=1}^{4}\frac{C_{j}(k) e^{i\delta\omega_{j}(k) t}}{\delta\omega_{j}(k)}
\\
\n&
\approx
\frac{-i C(k) a_{k_0}}{k^2\omega_{k_0}}e^{i\omega_{k_0}t},
\end{align}
where $C_{1...4}(k)$ and $C(k)=\sum C_i(k)$ are some complex
integration constants that we assume to have random phases with the
amplitude independent of $k$, hence, we can use that
$C(k)C^*(k)=\tilde{C}$.

Therefore, the non--resonant input term $\mathcal{N}^{nR}_{k_0}$ \cref{eq:nra}
depends on $a_{k_0}$ and $t$ as
\begin{align}
\label{eq:nraa}
\mathcal{N}^{nR}_{k_0}
&\approx
\frac{2\tilde{C}}{3\omega_{k_0}^2 |k_0|^3} 
e^{-i\omega_{k_0}t}a_{k_0}a^*_{k_0},
\end{align}
where terms with $C(k)C(k)$ and its complex conjugate vanish because
of randomness of phases.

More accurate estimation of $\mathcal{N}^{nR}_{k_0}$ will require
evaluation of integrals similar to
\begin{align}
\label{eq:nrInt}
I^{nR} &= \int \frac{e^{-i\left(\delta\omega_{i_1}(k) -
    \delta\omega_{i_2}(k)-\delta\omega_{i_3}(k\pm k_0)\right)t}}{k^2(k\pm
  k_0)^2\delta\omega_{i_2}(k)\delta\omega_{i_3}(k\pm k_0)} dk\n &=
\frac{1}{\omega_{k_0}^2 |k_0|^3}\int\frac{e^{-i\omega_{k_0}(1-1/k\pm
    1/(k\pm 1))t}} {k^2(k\pm 1)^2(1-1/k\pm 1/(k\pm 1))^2} dk,
\end{align}
that although result in more complex expressions, nevertheless, have
the same $e^{-i\omega_{k_0} t}$ asymptotic behavior for
$t\rightarrow\infty$.

Therefore, an equation for the longest wave length brain mode
$a_{k_0}$ that integrates the nonlinear non--resonant input from
smaller spatial scales can be written as
\begin{align}
\label{eq:akz}
\der{a_{k_0}}{t} &= \frac{a_{k_0}}{k_0^2}\left(
\gamma  + 
\frac{\beta
e^{-i\omega_{k_0}t}
}{\Omega^2 |k_0|}
a^*_{k_0}
\right)
 - 2\alpha |a_{k_0}|a_{k_0},
\end{align}
where $\gamma$ describes the excitation strength and $\beta$ is the
strength of non--resonant coupling. The last term (with the parameter
$\alpha$) was included to ensure that coupling does not produce
an overall mean field excitation, as well as to ensure that
in the limit of vanishing coupling ($\beta=0$) the solution of
\cref{eq:akz}
\begin{align}
\label{eq:aksol}
a_{k_0}(t) = \frac{\gamma}
{C_0\gamma \exp(-\gamma t/k_{0}^2) + 2\alpha k_{0}^2},
\end{align}
(where $C_0$ is a constant) has the same $1/k_{0}^2$ asymptotic
behavior for $t\rightarrow \infty$ as the solution of
\cref{eq:phiTensor} obtained with time and space scale independent
forcing.

The equation \cref{eq:akz} can be converted to a system of equations
for the amplitude $A$ and phase $B$ ($a_{k_0}=A e^{iB}$) as
\begin{align}
\label{eq:A0}
\der{A}{t} &= \frac{A}{k_0^2}\left[
\gamma + 
\frac{\beta A
\cos(B+\omega_{k_0}t-\delta_A)
}{\Omega^2 |k_0|}
\right]
 - 2\alpha A^2,
\\
\label{eq:B0}
\der{B}{t} &= -\frac{\beta A}{\Omega^2 |k_0|^3} \sin(B+\omega_{k_0}t-\delta_B),
\end{align}
where $\delta_A$ and $\delta_B$ were added to introduce tunable phase
delays \citep{2004PhRvL..93q4102A}. The system of equations
\cref{eq:A0} and \cref{eq:B0} includes non--resonant input from all
wave modes, but the resonant term \cref{eq:Nko} should also be added
together with additional equations for resonant wave amplitudes that
participate in resonant coupling. We will consider this complete
system later, but first we will investigate the behavior of the
nonlinear non--resonant part only.

\begin{figure}[!tbh] \centering
\includegraphics[width=1.0\columnwidth]{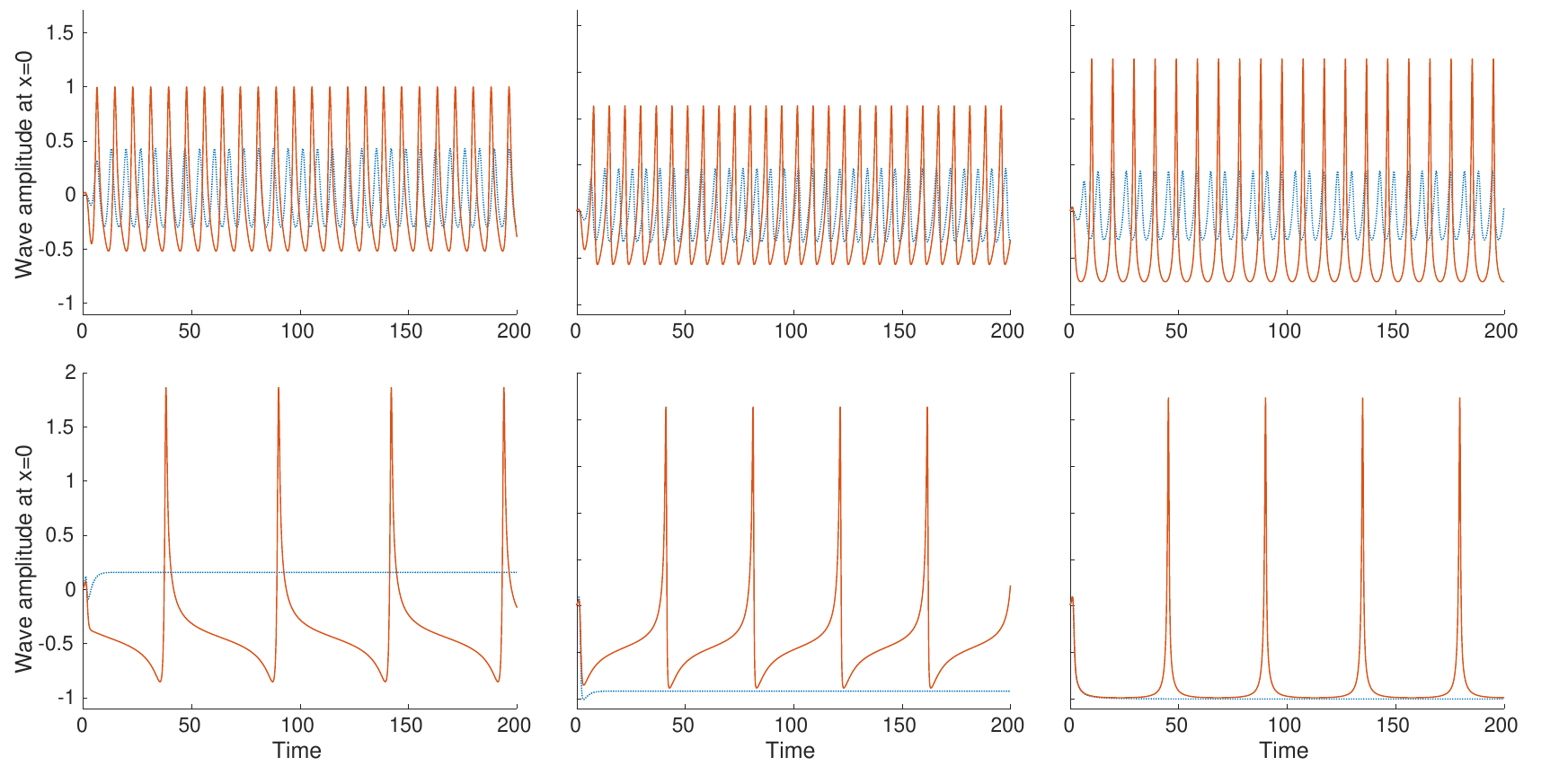}
\caption[] {The results of numerical integration of the system
  \cref{eq:A0,eq:B0}, that is time evolution of potential $\phi(x,t)$
  at $x=0$ or $A(t)\cos(B(t)+\omega_{k_0}t)$. For all plots the values
  of $\omega_{k_0}$, $k_0$, $\alpha$ and $\beta$ were set to be equal
  to 1, $\delta_A=0$, and $\gamma$ and $\delta_B$ were varied. The
  right,middle and left columns show plots for phase delay $\delta_B$
  equals to $\pi/4$, $3\pi/4$ and $\pi/2$ respectively. The top row
  displays transformation from weakly nonlinear oscillations shown by
  blue dotted lines for $\gamma=0.75$ to more strongly nonlinear
  regime (solid line, $\gamma=1.5$ (left and middle) and 2.25
  (right)). The bottom row shows the strongest nonlinear spiking--like
  time evolution of potential $\phi$ (solid line, $\gamma=2.55$ (left
  and middle) and 2.96 (right)) and its transformation to
  non-oscillatory (blue dotted line) regime for $\gamma=3$ (time and
  amplitude units are arbitrary).
\label{fig:1} }
\end{figure}

\Cref{fig:1} shows the results of numerical solution of the system
\cref{eq:A0,eq:B0} for several different sets of parameters. The time
evolution of highest frequency, longest wave length mode exhibits a
variety of types of oscillatory behavior, ranging from slightly
nonlinear modified sinusoidal shapes to more nonlinear looking shapes
similar to network attributed alpha waves or $\mu$-shaped oscillations
\citep{buzsaki2006rhythms}. Increase in the level of activation
$\gamma$ produces nonlinear signal with spike--like shape of a single
neuron firing.

It is interesting that this spiking--like solution of system
\cref{eq:A0,eq:B0} appears near the critical point, the oscillatory
state undergoes bifurcation and transitions to non-oscillatory regime
as $\gamma$ reaches the value above some critical point. To illustrate
the reason for this transition we will consider the simplest case of
$\delta_A=0$ and $\delta_B=\pi/2$ (although different $\delta_{A,B}$
values can be used for a similar analysis as well). The
non--oscillatory regime can be reached if $d A/dt \rightarrow 0$ and
$d B/dt \rightarrow -\omega_{k_0}$ as $t\rightarrow\infty$. Then from
\cref{eq:A0,eq:B0} one can write that at $t\rightarrow\infty$
\begin{align}
\gamma A - \omega_{k_0}A - 2 \alpha A^2 = 0,\qquad
\beta A \cos B_0 = -\omega_{k_0},\nonumber
\end{align}
where $B_0$ is some arbitrary constant phase.
Therefore, the non--oscillatory state requires that $\gamma$ satisfies to
\begin{align}
\gamma = \omega_{k_0} \left(1-\frac{2\alpha}{\beta \cos B_0}\right).
\end{align}
Hence for
\begin{align}
\omega_{k_0} \left(1-\frac{2\alpha}{\beta}\right) < 
\gamma < \omega_{k_0} \left(1+\frac{2\alpha}{\beta}\right),
\end{align}
the non-oscillatory solution is not possible. The simulations shown in
the bottom panels of \Cref{fig:1} confirm that the critical $\gamma$
value is indeed 3 when $\omega_{k_0}=\alpha=\beta=1$. Similar analysis
when $\delta_A=\delta_B$ gives the critical $\gamma$ value equals to
$(2-\cos{(\pi/3)})/\sin{(\pi/3)} \approx 1.732$.

\begin{figure}[!tbh] \centering
\includegraphics[width=0.9\columnwidth]{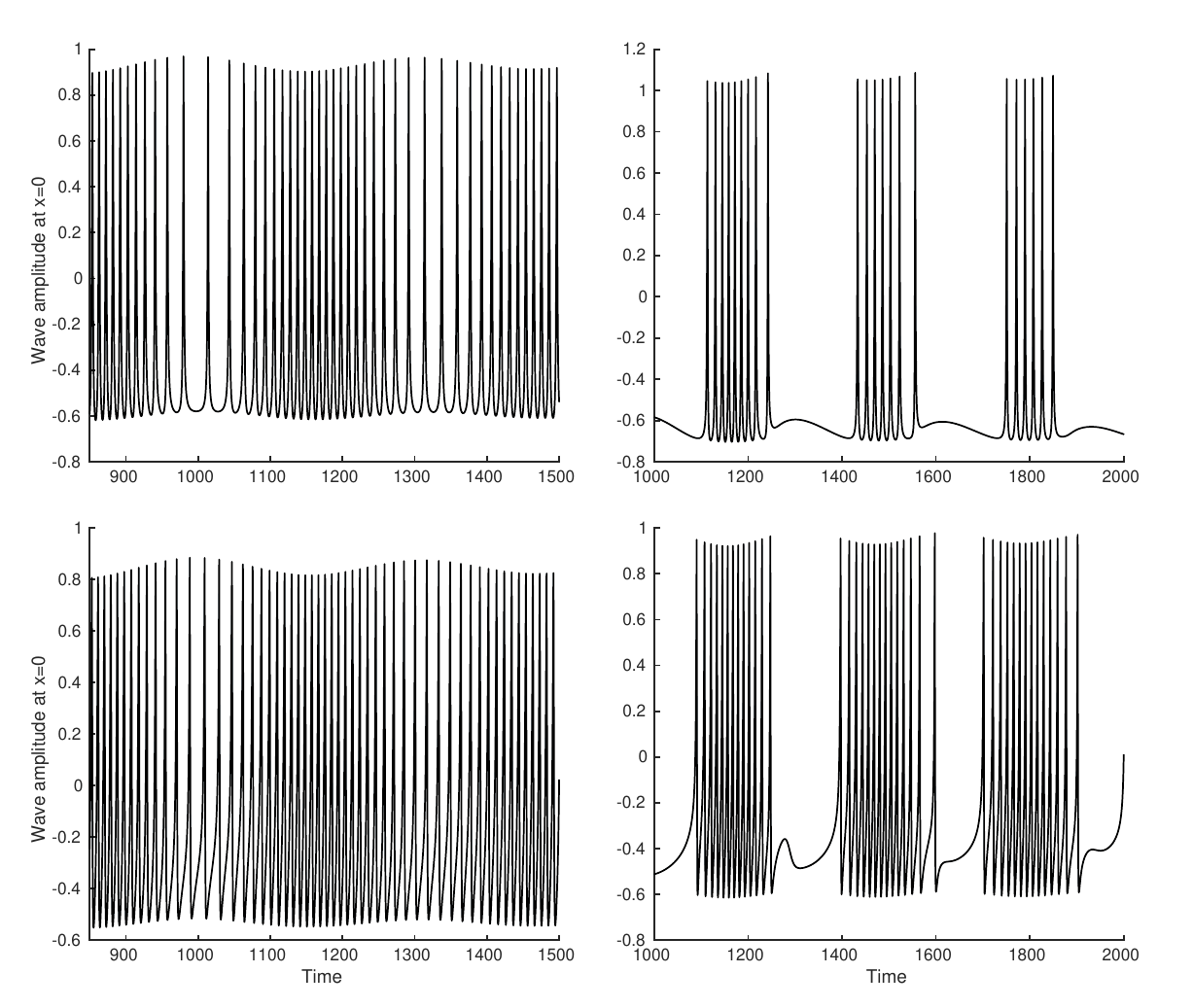}
\caption[] { The results of numerical integration of the system
  \cref{eq:A0,eq:B0} when exponential term was replaced by $I^{nR}$
  integrals \cref{eq:nrInt} with the region of integration set to $50
  k_0 < k < 1000 k_0$. For all plots the values of $\omega_{k_0}$,
  $k_0$, $\alpha$ and $\beta$ were set to be equal to 1, $\delta_A=0$,
  and $\gamma$ and $\delta_B$ were varied. The top and bottom rows
  show plots for phase delay $\delta_B$ equals to $3\pi/4$ and $\pi/2$
  respectively. The left column displays modulation of spiking rate
  for $\gamma=4.5$. The right column shows the nonlinear bursting of
  spikes for $\gamma=5.1$ (time and amplitude units are arbitrary).
\label{fig:2}}
\end{figure}

We would like to emphasize that all variety of models used for a
description of action potential neuron spikes, starting from the
seminal model by Hodgkin and Huxley \citep{pmid12991237}, and
finishing with many dynamical integrate--and--fire models of neuron
\citep{Gerstner:2014:NDS:2635959}, are based on an approximation of
several local neuron variables, e.g.~membrane currents, gate voltages,
etc., and defining the relations between these local
properties. Contrary to this and rather unexpectedly, the equation
\cref{eq:akz} is obtained through an integration of a large number of
oscillatory brain wave modes non-resonantly interacting in
inhomogeneous anisotropic media and shows spiking pattern solutions
emerging as a result of this non--resonant multi--mode interaction
rather than as a consequence of empirical fitting of nonlinear model
to several locally measured parameters.  It is also important that the
equation \cref{eq:akz} can not be separated into ``fast'' and ``slow''
parts as typically required for functioning of ``traditional'' neuron
models.  Because of this we would like to reiterate that this equation
should not be viewed as a single neuron model and should not be
considered as an alternative to any of the single neuron models
\citep{pmid19431309,*Nagumo1962,*pmid7260316}. It describes a
mechanism for generation of synchronous spiking activity as a result
of a collective input from many non-resonant wave modes.  The
transition to the synchronous spiking activity occurs in the vicinity
of the critical point where a bifurcation from oscillatory to
non-oscillatory state happens, thus indirectly supporting the
sub-criticality hypothesis \citep{pmid29899335} of brain operation.

\begin{figure}[!tbh] \centering
\includegraphics[width=0.95\columnwidth]{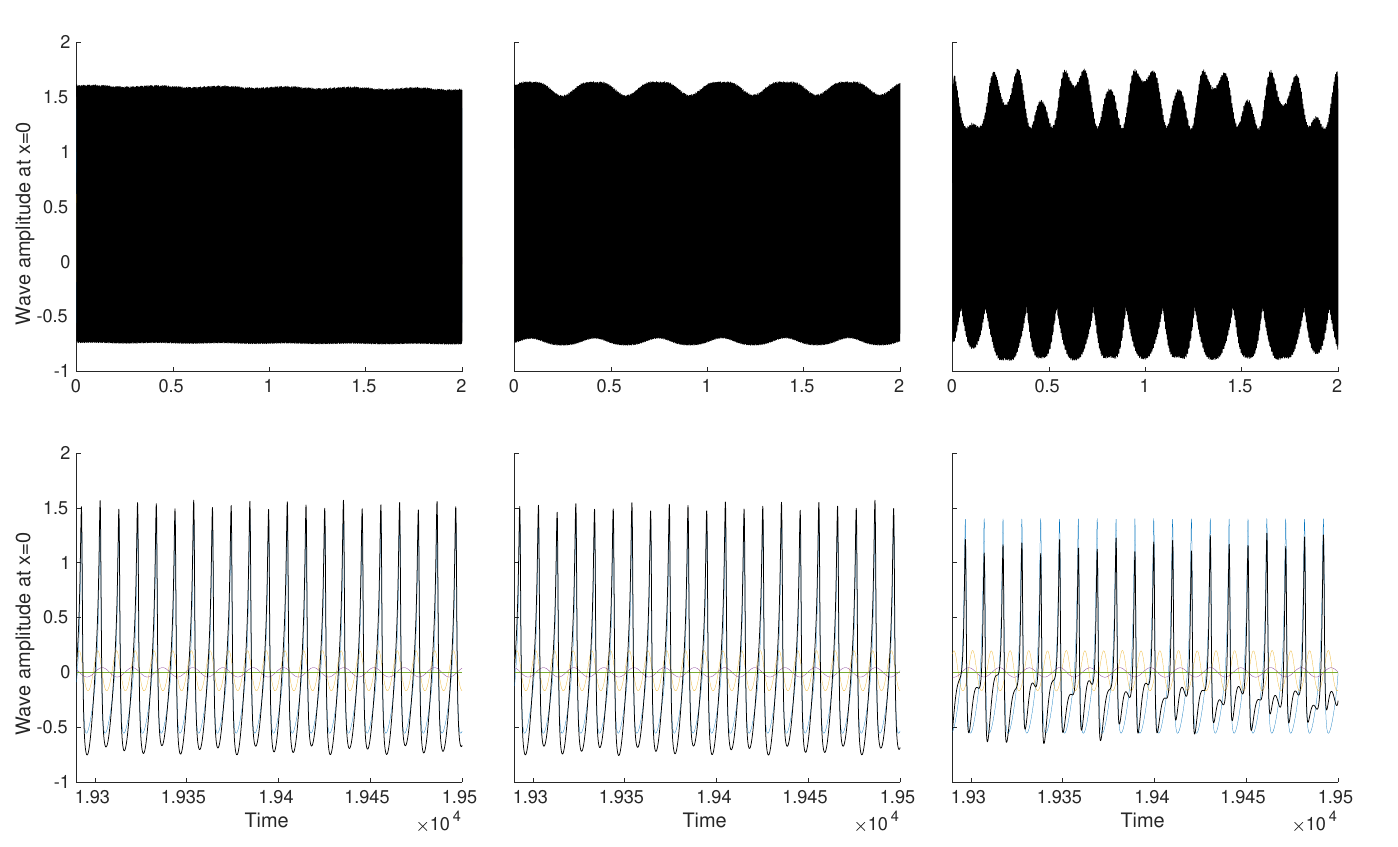}
\caption[]{
The results of numerical integration of the system \cref{eq:aknrr} for different values of weak resonant coupling $\lambda=0.001,0.01, 0.05$ (left, middle and right columns respectively). For all plots the values of $\omega_{k_0}$, $k_0$, $\alpha$ and $\beta$ were set to be equal to 1, and $\delta_A=\delta_B=\delta=3\pi/4$. The value of $\gamma$ is 1.535, that is sufficiently far from the criticality, but nevertheless large enough to modify an effective period for $k_0$ mode to be close to that of $k_1$. The total potential $\phi$ is plotted with the black and different colors show the oscillations of the individual modes. All plots clearly show emergence of low-frequency component as a result of increase of weak resonant coupling (time and amplitude units are arbitrary).
\label{fig:3}}
\end{figure}

As a next step we employed more complex expression for the total input from the non-resonant terms by including a sum of all $I^{nR}$ integrals \cref{eq:nrInt} instead of a single $e^{-i\omega_{k_0}t}$ exponent input. \Cref{fig:2} shows simulation results for several parameter sets with the same values as were used for plots of \cref{fig:1} ($\omega_{k_0}=\beta=k_0=1$, $\delta_A=0$). The numerical solution shows more complex behavior that includes now modulation of spiking rate with lower frequency and emergence of burst--like train of spikes, effects often observed in different types of neuronal activity \citep{Gerstner:2014:NDS:2635959}.

And finally, we considered a model that combines a chain of inputs from nonlinear modes generated due to resonant terms \cref{eq:Nko} into a set of non-resonant mode equations \cref{eq:akz}, that results in a system of equations for mode amplitudes $a_k$ for $k=k_0\dots k_N$ 
\begin{align}
\label{eq:aknrr}
\der{a_{k_0}}{t} &= 
\frac{a_{k_0}}{k_0^2}\left[
\gamma  + 
\frac{\beta}{\Omega |k_0|}
e^{-i\omega_{k_0}t+\delta}a^*_{k_0}
\right] 
- 2\alpha|a_{k_0}| a_{k_0} \n 
&+
\frac{\lambda}{k_0^2} a^*_{k_1}a_{k_2}
,\n
\der{a_{k_1}}{t} &= 
\frac{a_{k_1}}{k_1^2}\left[
\gamma  + 
\frac{\beta}{\Omega |k_1|}
e^{-i\omega_{k_1}t+\delta}a^*_{k_1}
\right] 
- 2\alpha|a_{k_1}| a_{k_1} \n 
&+
\frac{\lambda}{k_1^2} \left[ a_{k_0}a_{k_2} + a^*_{k_2}a_{k_3} \right]
,\n
& \dots \\
\der{a_{k_{n}}}{t} &= 
\frac{a_{k_n}}{k_n^2}\left[
\gamma  + 
\frac{\beta}{\Omega |k_n|}
e^{-i\omega_{k_n}t+\delta}a^*_{k_n}
\right] 
- 2\alpha|a_{k_n}| a_{k_n} \n 
&+
\frac{\lambda}{k_{n}^2} \left[ 
a_{k_{n-2}}a^*_{k_{n-1}} + 
a_{k_{n-1}}a_{k_{n+1}} + 
a^*_{k_{n+1}}a_{k_{n+2}}
\right]
,\n
& \dots \n
\der{a_{k_N}}{t} &= 
\frac{a_{k_N}}{k_N^2}\left[
\gamma  + 
\frac{\beta}{\Omega |k_N|}
e^{-i\omega_{k_N}t+\delta}a^*_{k_N}
\right] 
- 2\alpha|a_{k_N}| a_{k_N} \n 
&+
\frac{\lambda}{k_N^2} a_{k_{N-2}}a^*_{k_{N-1}}
,\nonumber
\end{align}
where the parameter $\lambda$ describes the strength of resonant
coupling between modes.

We would like to mention two new, rather important, and not entirely
obvious features that appear in the nonlinear system \cref{eq:aknrr},
but absent for phase coupled oscillators
\citep{Kuramoto2002,*Kuramoto2003}. First, the system \cref{eq:aknrr}
may show multiple critical point transitions corresponding to multiple
linear resonant frequencies $\omega_{k_i}$ as activation level
$\gamma$ increases. Second, sufficiently close to the critical point
the strong modification of an effective wave mode frequency by the
non--resonant input from multiple wave modes may result in nonlinear
resonances with different modes that are not possible for linear waves,
thus providing a 
mechanism for emergence of unexpected oscillations difficult to
explain by more simplistic models.

\begin{figure}[!tbh] \centering
\includegraphics[width=0.95\columnwidth]{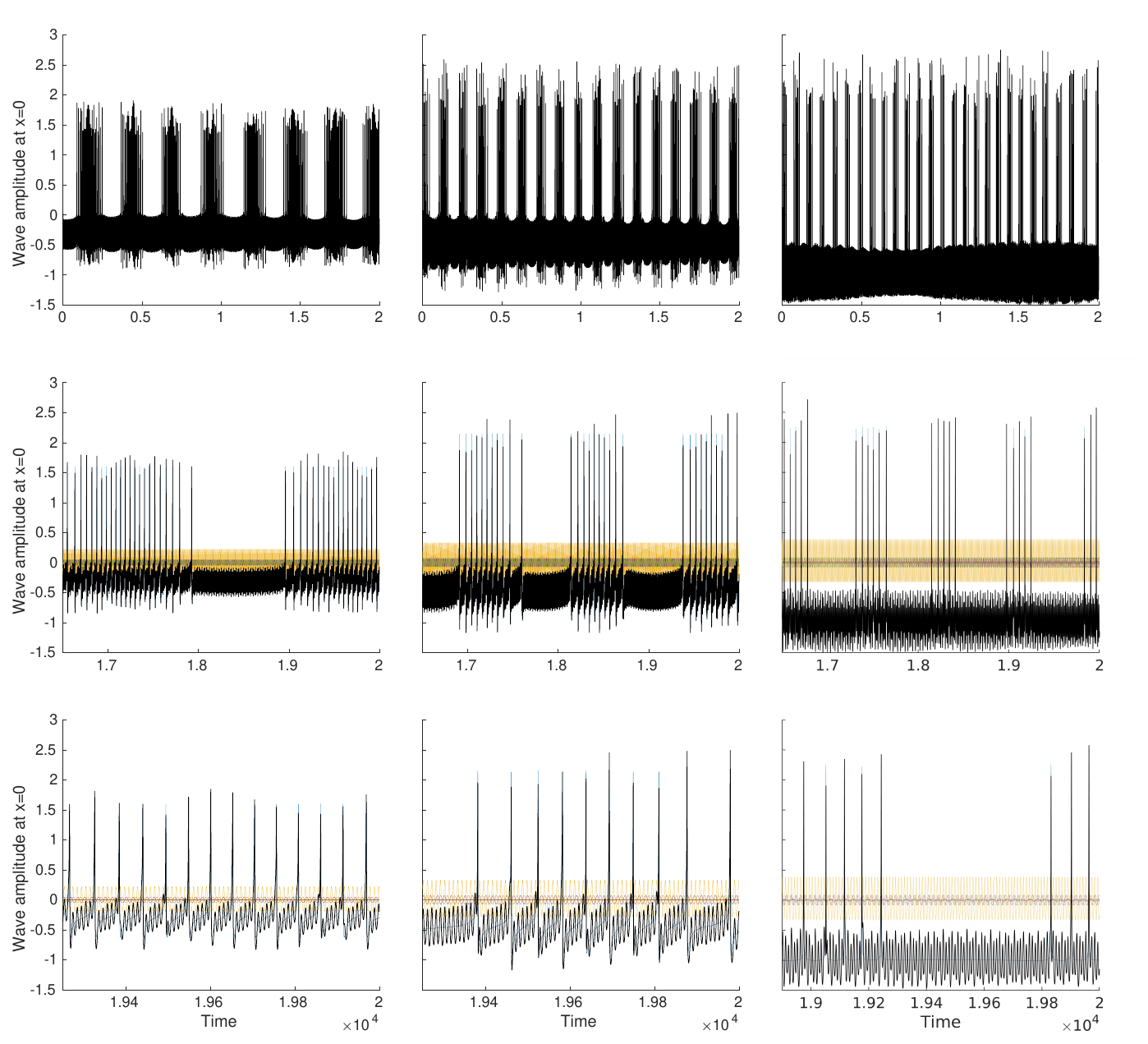}
\caption[]{ The results of numerical integration of the system
  \cref{eq:aknrr}. For all plots the values of $\omega_{k_0}$, $k_0$,
  $\alpha$ and $\beta$ were set to be equal to 1 and the resonant
  coupling $\lambda$ was 0.05. Different values of $\delta$ were again
  used in real and imaginary parts (as in \cref{eq:A0,eq:B0}) with
  $\delta_A=3\pi/4,0,0$, $\delta_B=3\pi/4,3\pi/4,\pi/2$ and close to
  the critical values of $\gamma=1.731, 2.575, 2,9969$ for the left,
  middle and right columns respectively. The total potential $\phi$ is
  plotted with the black and different colors show the oscillations of
  the individual modes. All plots show that when $\gamma$ is
  sufficiently close to criticality a week coupling produces jumps
  from subcritical to supercritical regimes with amazingly regular
  low--frequency quasiperiodicity (time and amplitude units are
  arbitrary).
\label{fig:4}}
\end{figure}

\Cref{fig:3,fig:4} show results of numerical simulation of the system
\cref{eq:aknrr}, clearly indicating that weak nonlinear resonant
coupling between just three modes with frequencies of $\omega_{k_0}$,
$2\omega_{k_0}/(1+\sqrt{5})$ and $2\omega_{k_0}/(3+\sqrt{5})$ is
capable of explaining an emergence of periodic activity with
frequencies up to 100-1000 times lower then the linear frequencies of
participating modes. We would like to emphasize again that the system
\cref{eq:aknrr} can not be separated into traditional
``slow'' and ``fast'' subsystems, hence the low frequency component
can not be explained by a modulation \citep{Rinzel1987} of the
``fast'' subsystem with oscillations of the ``slow'' part.

In \cref{fig:3} the high frequency spiking is generated with the level
of activation $\gamma=1.535$. This activation level is yet relatively
far from criticality but produces spikes with an effective rate
that is close to the next linear resonance frequency. The first,
second and third columns clearly show that small increase of the
resonant coupling (0.001, 0.01 and 0.05 respectively) results in
appearance of component with significantly lower frequency.

\Cref{fig:4} shows several simulations with the level of activation
that is close to criticality for the selected set of
parameters in each column. The small resonant coupling $\lambda=0.05$
in this case results in more profound effect of quasiperiodic shift of
oscillations back and forth from subcritical to supercritical regimes
effectively turning spiking on and off with low frequency.  Prediction
of the actual period of nonlinear low frequency oscillations
from the model parameters, e.g.~distances from the critical points and
other resonances, phase delays, etc., 
is an interesting open question that will be considered in future work.

In conclusion, in this paper we used a brain wave model \citep{bwl}
obtained from the general electrostatic form of Maxwell equations in
anisotropic and inhomogeneous media for description of interface
waves, i.e.~waves that propagate in the presence of interfaces
(surfaces, boundaries, membranes, transition regions, etc.), to show
that the unusual dispersion properties of those waves provide a
universal physical mechanism for emergence of low frequencies from
high frequency oscillations. The simple quadratic nonlinearity
introduced as a coupling source for the wave model allowed
the derivation of an equation for a nonlinear form
of those waves by taking a limit for a large number of non-resonantly
interacting wave modes, which we
emphasize is a limit that exists only due to unusual dispersion
properties of the waves.  The collective input from those
non--resonant modes results in nonlinear spiking--like solutions of
this equation and an existence of a bifurcation point from oscillatory
to non--oscillatory regime.  The multi--mode nonlinear system that
includes both non-resonant and resonant coupling between multiple
modes shows emergence of low frequency modulations as well as strongly
nonlinear low frequency quasiperiodic oscillations from subcritical to
supercritical regimes. This theory thus provides a basis for
relating quantitative tissue microstructural properties (such as
anisotropy and inhomogeneity) and measurable larger scale
architectural features (e.g, cortical thickness) directly to
electrophysiological measurements being performed increasingly
sensitive techniques (such as EEG) within a wide range of important
basic and clinical research programs.

\begin{acknowledgments}
LRF and VLG were supported by NSF grants DBI-1143389,
DBI-1147260, EF-0850369, PHY-1201238, ACI-1440412, ACI-1550405 and
NIH grant R01 MH096100. 
\end{acknowledgments}
\vspace*{-10pt}
%

\end{document}